\documentclass[12pt,preprint]{aastex}

\slugcomment{Submitted to the Astrophysical Journal Letters.}

\shorttitle{Relativistic flows in the hotspots}
\shortauthors{Georganopoulos \& Kazanas}

\begin{document}

\title{Relativistic and slowing down: the flow in the hotspots 
of powerful radio galaxies and quasars.}

\author{Markos Georganopoulos\altaffilmark{1}
 \& Demosthenes Kazanas\altaffilmark{2}}
\affil{Laboratory for High Energy Astrophysics, NASA Goddard Space Flight Center, 
Code 661, Greenbelt, MD 20771, USA.}
\altaffiltext{1}{Also NAS/NRC Research Associate; email:
markos@milkyway.gsfc.nasa.gov}
\altaffiltext{2}{email: Demos.Kazanas-1@nasa.gov}


\begin{abstract}
Pairs of radio emitting jets with lengths up to several hundred kiloparsecs 
emanate from the central region (the `core') of radio loud active galaxies.
 In the most powerful of them, these jets terminate in the `hotspots', 
compact high  brightness regions, where the jet flow collides with the 
intergalactic medium  (IGM).  Although it has long been established that in
 their inner  ($\sim$parsec) regions  these jet flows are relativistic,
it is still not
clear if they remain so at their largest  (hundreds of kiloparsec) scales. 
We argue that the X-ray, optical and radio data of  the hotspots, despite
 their at-first-sight disparate properties, can be unified in a 
scheme involving a relativistic flow upstream of the hotspot that
 decelerates to the  sub-relativistic speed of its inferred advance through
 the IGM and viewed at  different angles to its direction of motion. 
This scheme, besides providing an  account of the hotspot spectral 
properties with jet orientation, it also suggests that 
the large-scale jets remain relativistic all the way to the hotspots.  
\end{abstract}


\section{Introduction}

Radio maps of powerful radio galaxies usually show pairs of hotspots, compact 
synchrotron emitting regions, on each side of the radio core, 
feeding the extended lobes  of the system with radio emitting plasma. 
Cygnus-A, the prototype powerful radio  galaxy, was the first such source
whose hotspots were detected in X-rays too (Harris, Carilli \& Perley 1994); 
their radio--X-ray data were modeled successfully as synchrotron self-Compton  
emission from relativistic electrons in pressure equipartition with the
magnetic  field of the source (Synchrotron Self-Compton in Equipartition, SSCE). 
Equipartition, as yet without a firm theoretical justification, is favored  
energetically as it is the most economical way for producing a given 
synchrotron power. None of the two  Cygnus-A hotspots has been detected in the
optical, suggesting that the synchrotron  emission cuts-off at lower frequencies. 
Although {\it Chandra} observations of Cygnus-A (Wilson, Young \& Shopbell 2000)  
confirm the SSCE models, {\it Chandra} observations of Pictor-A (Wilson, Young 
\& Shopbell 2001), another powerful radio  galaxy, present a drastically different 
picture: While both its hotspots are seen in the  radio, X-rays reveal the 
presence of an one-sided jet  and a single hotspot on the jet side  of the 
source; furthermore, optical (presumably synchrotron) emission has been  
detected, but only from the hotspot seen in X-rays.  Additionally,  SSCE models 
strongly under-predict the  X-ray flux and imply that a magnetic field  $\sim 14$ times 
lower than the equipartition value is required to reproduce the X-ray flux.

These differences between the hotspot properties of Cygnus-A and Pictor-A are
in fact not incidental but rather representative of the radio galaxy 
subclass to which they belong (Cygnus-A is a NLRG -- Narrow Line Radio 
Galaxy, while Pictor-A a BLRG -- Broad Line Radio Galaxy; see also \S 2)
due, in part, to source orientation. According to the unification 
scheme of powerful radio galaxies and quasars (e.g. Urry \& Padovani 
1995), the Broad Line Region is only visible in objects with  jets 
axes relatively close to the observer's line of sight, while 
the  Narrow Line Region is visible at all inclinations.
Their difference in orientation is further supported by an altogether 
independent indicator, the ratio of the core-to-extended (lobe) radio 
power $R$. This is larger for more aligned objects,  because the lobe 
emission is isotropic, while the core emission is relativistically beamed 
in the direction of the plasma flow. 
The drastically different  values of $R$ between Cygnus-A  ($\log R\approx-3.3 
$; Zirbel \& Baum 1995) and Pictor-A ($\log R\approx-1.2$; Zierbel \& Baum 
1995), then, argue  that the jet axis of Cygnus-A is  closer to the plane of 
the sky, whereas that of Pictor-A is pointing closer to the observer's 
line of sight, a point further corroborated by the presence of the one-sided
X-ray jet in the same direction with the known VLBI jet  (Tingay et al. 2000)
of the core of  Pictor-A. The possibility of two intrinsically  asymmetric
 jets feeding the hotspots of Pictor A
is not favored,  given the practically equal power radiated by
the two extended radio lobes (Perley, R\"oser \& Meisenheimer 1997). 

In this note we put forward a hypothesis that provides an 
economical description of the multi-frequency, multi-object hotspot
data, which we summarize in \S 2. In \S 3 we describe our proposal 
and in \S 4 conclusions are drawn and some of the main open issues are 
discussed.

\section{Collective X-ray--Detected Hotspot Properties}

Table 1  lists all radio galaxies and quasars
with detected hotspot X-ray emission. The strong correlation 
of  X-ray detections from a single only or both hotspots with the 
jet orientation indicators (values of $R$, BLRG-NLRG classification) 
is unmistakable and argues, convincingly in our view, of the 
importance of orientation in the hotspot X-ray morphology.
Also apparent in Table 1 is the correlation between jet orientation
and the hotspot spectral properties and modeling: (a) In NLRG X-ray 
emission (generally attributed to the inverse  Compton process) is observed 
from both hotspots [except for the peculiar NLRG 3C 123 (Hardcastle, Birkinshaw 
\& Worall 2001) which exhibits detected X-rays from only one hotspot] and
it is successfully modeled as SSCE emission. Also, NLRG show no optical 
hotspot emission, suggestive of a cut-off in the observed synchrotron 
spectrum at lower frequencies [the optical emission in the NLRG 3C 295 is 
modeled as self-Compton (Brunetti 2002b)]. (b) In BLRG and quasars, which 
generally exhibit one-sided VLBI jets (e.g. Giovannini et al. 2001), a 
single hotspot is detected in X-rays, always on the VLBI jet side; 
synchrotron optical emission is also  detected consistently from the 
same hotspot. Models of the 
radio -- X-ray spectra fail to produce the observed X-ray fluxes with 
SSCE parameters, severely under-predicting their observed X-ray fluxes. 
Forcing SSC models to produce X-ray fluxes in agreement with observation
leads to magnetic field values several times below the inferred 
equipartition ones. An exception 
to this rule is 3C 263, for which SSC models can reproduce the observed 
X-ray  flux with a magnetic field only half of its equipartition value 
(Hardcastle et al. 2002). 

Closer inspection of Table 1 reveals the following correlations between 
source orientation and photon frequency: 
($i$) {\sl X-ray} hotspot emission correlates with source alignment 
(as  measured by $R$): Sources closer to the plane of the sky (small $R$) 
exhibit X-ray emission of comparable flux from both hotspots. 
As alignment increases the X-ray hotspot emission is restricted 
to the near hotspot (i.e. the 
one on the same side as the VLBI  jet); furthermore, its X-ray-to-radio 
ratio becomes larger than predicted by SSCE, hinting the presence of an 
additional component more  sensitive to orientation effects than SSC. ($ii$) 
{\sl Optical} (synchrotron) emission, initially  weak or absent, appears 
at the near hotspot as $R$ gets larger. [The only known exception  to this 
sequence is possibly the quasar 3C 207 (Brunetti et al. 2002a), which 
has NLRG hotspot characteristics but exhibits broad line emission 
and strong core dominance ($\log R \approx -0.5$; Hough \& Readhead 
1989). However, VLBI observations (Hough et al. 2002) show that the 
parsec scale radio jet of this source is  strongly bent, exhibiting 
large-scale properties akin to a NLRG, while at small scales  appears 
to be a BLRG]. ($iii$) {\sl Radio} emission is consistently observed 
from both hotspots  in all cases. However, while sources with jet axis 
close to the plane of the sky exhibit  practically  equal hotspot fluxes, 
for more aligned jets the near hotspot has a higher radio  flux and a 
flatter radio spectrum (Dennett-Thorpe et al. 1997).

\section{Frequency-dependent beaming}

In this note we propose that the above properties can be accounted for 
by postulating relativistic beaming which becomes weaker (i.e. of  lower Lorentz 
factor, broader beaming angle) with decreasing photon frequency. 
Such a frequency dependent beaming arises naturally in the decelerating flow 
following the shock expected in the transition of  the jet velocity from 
relativistic to that of the hotspot advance [$ v_{adv}\approx 0.1 c$, 
(Arshakian \& Longair 2000)] through the IGM. Electrons, accelerated 
impulsively to relativistic energies  at the shock, are transported 
downstream with the fluid, while at the same time losing  energy to 
synchrotron and inverse Compton radiation. The highest energy electrons 
are  therefore located immediately downstream of the shock where the  
flow Lorentz factor is also higher.

This model provides a direct account of the 
properties of the synchrotron component: 
According to our proposal, beaming is  highest for the {\sl X-ray 
synchrotron} emission, consistent  with its (possible) 
detection from a single object, namely the superluminal 
 3C 390.3 (Prieto 1997).  The {\sl Optical synchrotron}  emission is still 
significantly  beamed (but less than the synchrotron X-rays) and it is 
(as expected) preferentially detected in the  hotspots of the approaching
jet in the more aligned objects (BLRG). The synchrotron emission from 
the receding jet hotspot is beamed away from the observer with its flux 
reduced by a factor $R_h=(1+\beta\cos\theta)^{2+\alpha}/(1-\beta\cos
\theta)^{2+\alpha}$, where  $\theta$ is the angle between the line of 
sight and the jet axis, $\beta = v/c$ is the velocity of the fluid and 
$\alpha$ is the  spectral index. The downstream velocity 
perpendicular to the shock front is $v=c/3$ in the hotspot frame, which, 
added to the  hotspot advance-velocity of $v_{adv}\approx 0.1\; c$, 
transforms to $v\approx 0.42\; c$  in the lab frame (assuming a 
perpendicular shock). Even for  this most restrictive case, the near-to-far
hotspot flux  ratio  is $R_h\approx 10 $   for $\alpha=1$ and  $\theta=30^o$,
an angle typical (Urry \& Padovani 1995) of BLRG and quasars.
However, this is only a lower limit, as oblique shocks, commonly seen in
numerical  simulations (Komissarov \& Falle 1996, Aloy et al. 1999), 
routinely produce  hotspot flows 
with Lorentz factors up to $\Gamma \approx 3-4$ that  decelerate to eventually 
match the sub-relativistic hotspot advance speed [this has been 
used by  Komissarov \& Falle (1996)  to explain the fact that the radio
hotspots on the side of the jet are systematically  brighter
and have a flatter spectrum]. Flows with $\Gamma\approx 2$ suffice to
suppress the optical flux from the far hotspot of a BLRG or a quasar 
by more than two orders of magnitude.  Finally, 
the effects of beaming are smallest for the {\sl Radio} emission, which is 
observed from the hotspots of both jets in all objects, albeit with
systematically higher intensities for the near jet hotspot.

The situation with the inverse Compton component, the generally 
accepted process responsible for the X-ray emission, is more subtle:
The detection of X-rays consistent with SSCE models from the hotspots 
of {\sl both} jets in NLRG, argues that this emission (like the 
radio) is not significantly beamed in these objects (beaming affects 
both these components identically within SSC). However, the increase 
of the X-ray flux relative to that of the radio with increasing $R$, 
suggests  contribution from a component with beaming enhancement 
more pronounced than that of SSC. The obvious candidate is inverse Compton 
scattering of photons other than  those locally produced by synchrotron 
(the so-called External Compton, EC), whose beaming properties are more 
sensitive (Dermer 1995; Georganopoulos, Kirk \& Mastichiadis 2001) 
to orientation  than 
those of SSC. In fact, this process has been invoked (Tavecchio et al. 
2000; Celotti, Ghisellini \& Chiaberge 2001) to account for the X-ray 
emission from the knots of the PKS 0637-752 jet by considering X-ray 
production from the up-scattering of the Cosmic Microwave Background 
(CMB) photons; however, the CMB photons alone are in general inadequate to 
account for the X-ray emission of most hotspots.

In search of an additional photon component necessary to produce the
observed enhancement in Comptonized flux with decreasing angle of 
observation, we have come to the realization that in a 
decelerating flow, synchrotron photons from its slower ($v \simeq 0.1 \, 
c$) parts, can serve as seed  photons for 
scattering by the electrons of its faster, likely relativistic, section. As
in EC scattering (Dermer 1995), the photon energy density in the comoving  frame of the fast electrons
 increases by $\Gamma_{rel}^2$, where $\Gamma_{rel}$ is the
relative Lorentz factor between the downstream slow plasma emitting the synchrotron
 seed photons and the upstream fast plasma containing the scattering electrons.
Because the density of these photons exceeds that of the CMB ones (for 
the parameters of most X-ray detected hotspots), they present the 
dominant soft photon source responsible for the X-ray emission. {\it It is 
our contention that  Compton scattering of  these  photons off the 
electrons of the upstream, faster flow (Upstream Compton, UC) can lead to the
necessary enhancement of the X-ray-relative-to-radio flux observed
in BLRG.}

We demonstrate the above features with a kinematic, one-dimensional 
model which  assumes the injection of an $\gamma^{-2}$ power law electron 
energy distribution with a high energy cut-off $\gamma_c = 2 
\cdot 10^6$ at the base of a relativistic,  decelerating flow. 
In Figure 1 we exhibit the spectrum resulting from such a flow 
viewed at two different angles.  The electron energy density is 
assumed to be in equipartition with a magnetic field of $B_{eq}=3
\cdot10^{-4}$ G, (similar to the observationally
determined equipartition field of Cygnus-A), while the position 
dependent flow Lorentz factor to have the form $\Gamma(z)=\Gamma_0
(z/z_0)^{-2}$  with $\Gamma_0=3$ and $z_0 \simeq 1$ kpc and similar
transverse dimension, as determined by the {\it Chandra} observations. 
The observed luminosity (the angularly resolved emission allows the
computation of the absolute luminosity scale) from the entire
volume of the flow (taking into account the $z-$ and $\theta-$dependent 
relativistic beaming) is calculated as a function of the photon energy,
for two different observing angles, $\theta=20^\circ$ (solid) and 
$\theta=70^\circ$ (dashed).

The Spectral Energy Distribution (SED) for $\theta= 70^\circ$ is 
very similar to those of the NLRG hotspots, in particular 
that of Cygnus-A. Specifically, one should note the synchrotron 
component cutoff below the optical band and the low  
X-ray-to-radio luminosity ratio. At $\theta=20^\circ$ the resulting
SED is very different, resembling that of the BLRG/quasar hotspots: 
Its synchrotron slope is flatter 
and peaks at optical energies. Also, the  ratio of the X-ray-to-radio 
luminosity is higher, consistent with the ratios observed for these
more aligned sources. 
A noteworthy feature of the spectra of figure 1 is the 
distinctly  different slopes of the synchrotron component at large and 
small values of $\theta$, in agreement with observations which require 
harder spectra from the near  compared to the far hotspots
(Dennett-Thorpe et al. 1997), a systematic never convincingly explained. 
Furthermore, the cooling break of $\Delta \alpha = 1/2$  
at radio frequencies for $\theta = 70^\circ$ (similar to that observed 
in the hotspots of Cygnus-A) is significantly reduced at $\theta = 
20^\circ$ and it can, with a judicious choice of parameters, even 
disappear (as is the case in Pictor-A).

The harder spectra obtained for  smaller $\theta$ and the reduction 
in the cooling break change in slope from its canonical 
value of $\Delta \alpha = 1/2$ can be easily understood if one 
considers that the IR-optical emission receives contribution from
 electrons of both higher energy and higher flow speed, i.e. of 
increased relativistic boosting. For small values of $\theta$ this 
amplifies the IR-optical emission relative to that of the radio, which 
is produced by electrons from the slower parts of the flow. This 
preferential relativistic boosting of the higher frequency radiation 
results, hence, in harder spectra and also smaller slope changes  
associated with the cooling break. At large $\theta$, the effects of 
differential boosting are negligible and one obtains the standard
values for the synchrotron slopes and cooling break.
Similarly, the X-ray emission at small $\theta$ is dominated by the more
beamed UC emission; this is Doppler-suppressed at larger $\theta$ and 
the dominant contribution to the observed X-ray flux  is due to the  
semi-isotropic SSCE emission from the low speed part of the flow.
We plan to discuss these issues in greater detail in an upcoming 
publication.

\section{Discussion}

We have presented above a general framework which encompasses
the broad band (radio-to-X-ray) spectral properties of the 
X-ray detected hotspots of powerful radio galaxies. Our thesis is 
that the presence of a decelerating, relativistic ($\Gamma_0 \sim  
3$) flow in the hotspots explains  the ensemble of their multi-frequency, 
multi-object data as  an (essentially) one-parameter family, namely 
the angle $\theta$. 
The deceleration of the flow is the ingredient necessary to produce 
the increase of the X-ray-to-radio flux ratio in the BLRG relative 
to the NLRG hotspot spectra, their most prominent difference. As explained,
this is due to the {\it Upstream Compton (UC)} process which enhances
the Compton emission more than that of the synchotron for small values 
of $\theta$. The precise dependence of the flow's Lorentz factor with 
distance does affect the model spectra, however, it does not change their 
qualitative features (i.e. harder spectra,  higher energy synchrotron 
cutoffs for sources with smaller $\theta$).
As far as we know, this is the first 
concrete (albeit indirect) evidence that the flows in these jets 
remain relativistic all the way to the hotspots. 

We believe that within the limited sample of objects of 
Table 1, our model fares quite well in providing a broad framework
which can address and systematize the at-first-hand disparate 
multi-frequency observations of the radio galaxy hotspots. Exceptions 
from the expected properties, even within Table 1, could be attributed
to idiosyncrasies of the specific sources. In fact, in view of 
the peculiarities of  hotspot flows produced in simulations todate 
(e.g. Aloy et al. 1999; these exhibit time dependent multiple oblique 
shocks, while the flows turn around to flow backward in the outer 
layers of the jet ``cocoon"), it is rather surprising that 
the simplified 1-D model used in our calculations can describe the collective
data as  well as it does. After all, the distribution of objects in 
angle (and by consequence their properties) is not bimodal, 
so objects with intermediate, or mixed properties should 
soon appear as the list of ojbects gets larger. 

We should also stress the particular importance of deep multi-frequency
observations which could detect the hotspot emission of the counterjet
(preferentially in BLRG), in the determination of the flow velocity
field at the hotspots. This can be achieved through measurements of 
the flux ratio $R_h$ at various frequencies and, assuming an orientation 
angle $\theta$, invert its  expression to obtain the flow velocity 
$\beta=(R_h^{1/(2+a)}-1)/((R_h^{1/(2+a)}+1)\cos\theta)$ as a function of 
photon frequency.  An upper limit, $\theta_{max}$, to the value of 
the angle $\theta$ can be also estimated by setting $\beta=1$: 
$\cos \theta_{max}=(R_h^{1/(2+a)}-1)/((R_h^{1/(2+a)}+1)$.
We believe that direct comparison with detailed models 
will then produce in addition to other parameters, also the dependence
of $\Gamma$ on the distance.

An enlargement of this list, through X-ray ({\it Chandra}), optical, 
and radio (VLA) observations of sub-arcsecond resolution,
in combination with detailed 
fits of the combined radio-to-X-ray spectra  will lead to
a more stringent testing (and possible rejection) of this 
model. In this respect, future  GLAST observations may be of great 
importance by providing additional constraints in the 10 MeV
- 10 GeV band, where the IC flux peaks (see figure 1). 
We have estimated that in some of the nearby ($z\lesssim 0.1$) 
sources, the hotspot high energy $\gamma-$ray 
flux will be both detectable and its position sufficiently well
determined to be distinguished  from the (potential) emission 
from the AGN ``core''.


\clearpage

\begin{deluxetable}{lcccccccc}
\tablecolumns{10}
\tablecaption{X-ray hotspots. \label{tbl1}}
\tablewidth{0pt}
\tablehead{
\colhead{SOURCE} & \colhead{TYPE} & z &  \colhead{$\log\rm R $}   &\colhead{OPTICAL} &
   \colhead{X-RAY}& \colhead{SSCE}} 
\startdata

\objectname[]{3C 330} & NLRG &0.549 & -3.5  \tablenotemark{a}& NO &  
YES\tablenotemark{b}$\phantom{_{g,h}}$, two sides& YES  \\

\objectname[]{Cygnus A} & NLRG & 0.057 & -3.3 \tablenotemark{c} & NO   &
YES\tablenotemark{d,e}$\phantom{_{g,h}}$, two sides & YES \\

\objectname[]{3C 295} & NLRG &0.461 & -2.7 \tablenotemark{c} & YES, SSC \tablenotemark{f}& YES\tablenotemark{g,h}$\phantom{_{g,h}}$,  two sides & YES \\

\objectname[]{3C 123} & NLRG & 0.218 & -1.9 \tablenotemark{i} & NO  & YES\tablenotemark{j}
$\phantom{_{g,}}$, one side$\phantom{s}$  & YES \\


 \objectname[]{3C 263} & Q  &0.656 & -1.0 \tablenotemark{k} & YES\tablenotemark{b}$\phantom{_{m,}}$, 
jet side & YES\tablenotemark{b}$\phantom{_{g,g}}$, jet side$\phantom{ss}$ & see text \\

 \objectname[]{3C 351} & Q  &0.371 & -1.9 \tablenotemark{l} & YES\tablenotemark{m,b}$\phantom{_{m,}}$, jet side & YES\tablenotemark{m,b}$\phantom{_{g,h}}$, jet side$\phantom{ss}$ &NO  \\


 \objectname[]{Pictor A} & BLRG & 0.035 &  -1.16 \tablenotemark{c} & YES\tablenotemark{n}$\phantom{_{m,}}$, jet side &  YES\tablenotemark{n}$\phantom{_{g,h}}$, jet side$\phantom{ss}$ & NO \\

 \objectname[]{3C 303} & Q  & 0.141 &  -0.73 \tablenotemark{c} & YES\tablenotemark{o}$\phantom{_{m,}}$, jet side  & YES\tablenotemark{p}$\phantom{_{g,h}}$, jet side$\phantom{ss}$ &  NO \\

\objectname[]{3C 390.3} & BLRG & 0.057 & -1.06 \tablenotemark{c} & YES\tablenotemark{o}$\phantom{_{m,}}$, jet side & YES\tablenotemark{q}$\phantom{_{g,h}}$, jet side$\phantom{ss}$ & NO \\


\objectname[]{3C 207 } & Q & 0.684 &-0.5 \tablenotemark{k}& NO  &YES\tablenotemark{r}
$\phantom{_{g,}}$, jet side$\phantom{ss}$  & YES   \\

\enddata

\tablerefs{
(a) Saikia \& Kulkarni 1994,  
(b) Hardcastle et al. 2002,
(c)  Zirbel \& Baum 1995, 
(d) Harris, Carilli \& Perley 1994,
(e) Wilson, Young \& Shopbell 2000,
(f) Brunetti 2002b,
(g) Harris et al. 2000,
(h)Brunetti et al. 2001a,
(i) Hardcastle et al. 1998,
(j) Hardcastle et al. 2001,
(k) Hough \& Readhead 1989, 
(l) Wills \& Browne 1986,
(m) Brunetti et al. 2001b,
(n) Wilson et al. 2001,
(o)  L\"ahteenm\"aki \& Valtaoja 1999,
(p) Kataoka et al. 2003,
(q) Prieto 1997,
(r) Brunetti et al. 2002a.}

\end{deluxetable}
\clearpage

\begin{figure}
\epsscale{0.8}
\plotone{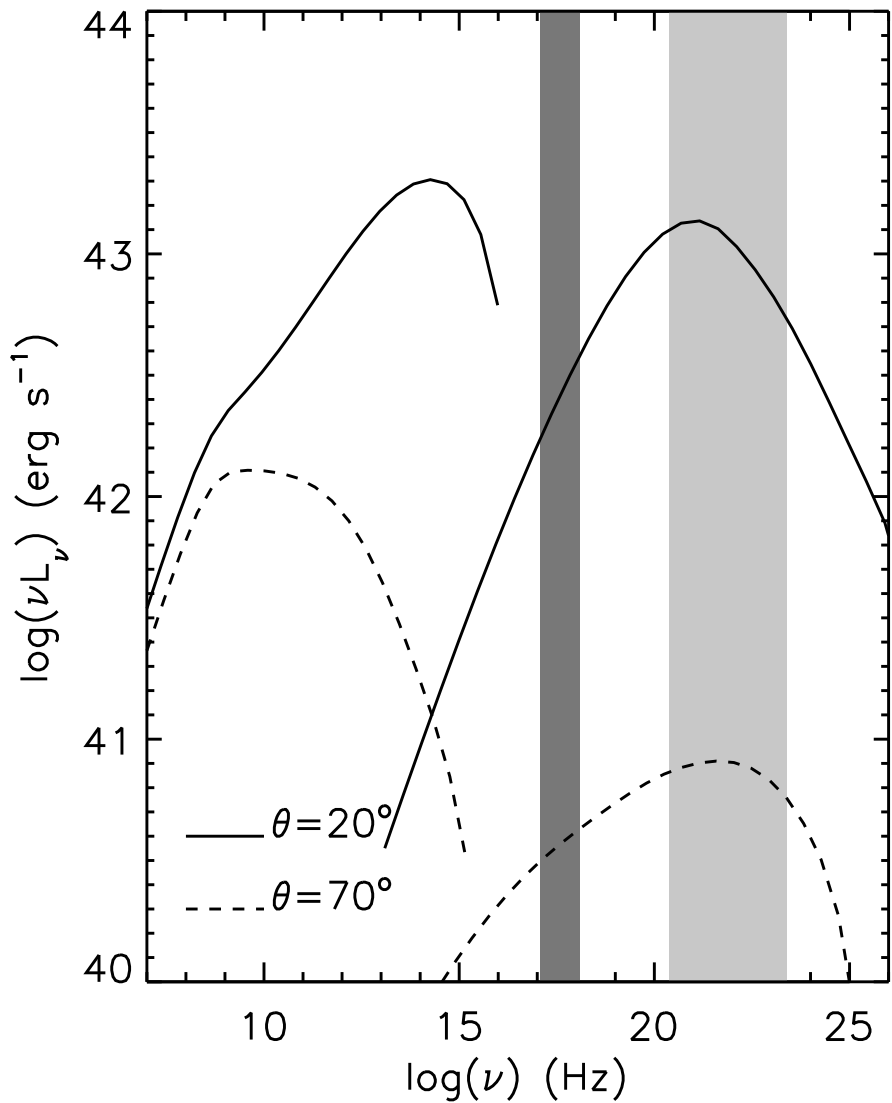}
\caption{The synchrotron-Compton models for the decelerating flow 
described in the text as viewed at two different angles, $\theta 
= 20^\circ$ (solid) and $\theta = 0^\circ$ (dashed). The dark gray
band denotes the {\it Chandra} band while the light one that of GLAST.
Note the different slopes of the synchrotron and Compton components
at the different angles and the increase of the radio-X-ray ratio
with decreasing $\theta$.}
\label{fig1}
\end{figure}

\end{document}